%% file: main.tex
\documentclass[aps,floatfix,twocolumn,amsmath,amssymb]{revtex4-2}
\usepackage{graphicx}
\usepackage{color}
\usepackage{bm}
\usepackage{hyperref}
\hypersetup{colorlinks=true,citecolor=blue,linkcolor=blue,urlcolor=blue}
\usepackage{newtxtext,newtxmath} 
\allowdisplaybreaks

\begin{document}

\title{%
Reentrance of the Disordered Phase
in the Antiferromagnetic Ising Model on a Square Lattice
with Longitudinal and Transverse Magnetic Fields
}

\author{Ryui Kaneko}
\email{rkaneko@phys.kindai.ac.jp}
\affiliation{%
Department of Physics, Kindai University, Higashi-Osaka, Osaka 577-8502, Japan}

\author{Yoshihide Douda}
\affiliation{%
Department of Physics, Kindai University, Higashi-Osaka, Osaka 577-8502, Japan}

\author{Shimpei Goto}
\email{goto.las@tmd.ac.jp}
\altaffiliation[Present address: ]{%
College of Liberal Arts and Sciences, Tokyo Medical and Dental University,
Ichikawa, Chiba 272-0827, Japan}
\affiliation{%
Department of Physics, Kindai University, Higashi-Osaka, Osaka 577-8502, Japan}

\author{Ippei Danshita}
\email{danshita@phys.kindai.ac.jp}
\affiliation{%
Department of Physics, Kindai University, Higashi-Osaka, Osaka 577-8502, Japan}

\date{\today}
 
\begin{abstract}
Motivated by recent experiments
with Rydberg atoms in an optical tweezer array,
we accurately map out the ground-state
phase diagram of the antiferromagnetic Ising model
on a square lattice
with longitudinal and transverse magnetic fields 
using the quantum Monte Carlo method.
For a small but nonzero transverse field,
the transition longitudinal field is found to
remain nearly constant.
By scrutinizing the phase diagram,
we uncover a narrow
region where the system exhibits reentrant transitions
between the disordered and antiferromagnetic phases
with increasing transverse field.
Our phase diagram provides a useful benchmark
for quantum simulation of a Rydberg atom system.
\end{abstract}


\maketitle



Quantum effects in many-body systems have been the subject of intensive
research.
Accurate simulation of quantum systems can reveal novel phases and
phase transitions.
Although numerical simulation on a classical computer is useful,
the number of tractable models is limited
because of the exponential growth of the Hilbert space.
An alternative approach is to use highly controllable devices,
namely, analog quantum simulators,
to emulate quantum many-body
systems~\cite{feynman1982,buluta2009,georgescu2014}.

Quantum simulators using Rydberg atoms in an optical tweezer array
have attracted growing interest owing to rapid technological
advances~\cite{browaeys2020}.
Optical tweezers allow one to hold and move each atom precisely.
In addition,
the distance between atoms is large enough that
individual atoms can be observed.
Because dipole--dipole interactions between Rydberg atoms are much stronger
than those between ground-state atoms, one can conduct experiments at
relatively high temperatures without evaporative cooling, and the
typical time scale of the real-time dynamics is roughly $1000$ times faster
than that of ultracold atoms in optical lattices.

Owing to these advantages, recent experiments using 
Rydberg atom arrays~\cite{bernien2017,keesling2019,
guardado-sanchez2018,lienhard2018,scholl2020,ebadi2020,
bluvstein2021,deleseleuc2019}
have successfully observed various interesting many-body phenomena.
For example, quantum phase transitions
and nonequilibrium dynamics~\cite{bernien2017,keesling2019}
have been observed in
Rydberg systems that realize
the one-dimensional Ising model
with longitudinal and transverse magnetic fields.
Furthermore,
symmetry-protected topological phases have been identified
in a simulator that imitates
the Su--Schrieffer--Heeger chain~\cite{deleseleuc2019}.
The simulation of not only systems in one spatial dimension
but also those in two spatial dimensions is
feasible~\cite{guardado-sanchez2018,lienhard2018}.
Very recently, the number of controllable atoms exceeded
$200$~\cite{scholl2020,ebadi2020}.

The recent development of quantum simulation experiments
has motivated a revival of
theoretical research on fundamental quantum spin models.
In particular,
the study of nonequilibrium dynamics is among the
most active fields.
For instance,
the observation of certain states that exhibit anomalously
slow thermalization~\cite{bernien2017,bluvstein2021}
has stimulated research on quantum many-body
scars~\cite{turner2018a,turner2018b,james2019,shibata2020}.
There are many open questions
on how quantum information propagates
in terms of the real-time dynamics of the quantum Ising
model~\cite{guardado-sanchez2018,lienhard2018}.

It is essential to understand the
ground-state properties of static systems
before tackling these unresolved problems.
The Ising model has served as a textbook example
of how to describe a phase transition
in statistical physics~\cite{sachdev2011,suzuki2013}
because of its simplicity and solvability~\cite{pfeuty1970}.
Rydberg systems are suitable
for realizing the quantum Ising model.
In these systems, the longitudinal and transverse fields
can be controlled by
frequency detuning and the Rabi frequency of the laser,
respectively~\cite{robicheaux2005}.
The ground-state phase diagrams of 
quantum Ising models on several lattices
have been extensively studied using
the quantum Monte Carlo (QMC)
method~\cite{bloete2002,moessner2000,moessner2001}.

Although many Ising models
have been analyzed,
the precise ground-state properties of the mixed-field Ising model
have yet to be explored on the simple square lattice.
The model is so primitive that
detailed analysis has been overlooked.
In one spatial dimension,
the precise phase diagram of the mixed-field Ising model
is determined by
the exact diagonalization
(ED)~\cite{sen2000,bonfim2019},
QMC~\cite{novotny1986}, and
density matrix renormalization group~\cite{ovchinnikov2003}
methods.
By contrast, in two spatial dimensions,
only a schematic phase diagram
for a few dozen sites has been drawn
in a recent ED study~\cite{lienhard2018}.


In this letter,
we draw the ground-state phase diagram
of the antiferromagnetic Ising model
on a square lattice
with both longitudinal and transverse fields.
The Hamiltonian of the mixed-field Ising model is defined as
\begin{equation}
 H
 = J \sum_{\langle i,j \rangle} \hat{S}_i^z \hat{S}_j^z
 - h \sum_{i} \hat{S}_i^z
 - \Gamma \sum_{i} \hat{S}_i^x,
\end{equation}
where $J(>0)$ denotes the strength of the
antiferromagnetic Ising interaction,
and $h$ ($\Gamma$) represents the longitudinal
(transverse) magnetic field.
The operators $\hat{S}^{z}$ and $\hat{S}^{x}$ are
the $z$- and $x$-component $S=1/2$ Pauli spin operators.
The notation $\langle i,j \rangle$ indicates that sites
$i$ and $j$ are nearest neighbors.
We set $\hbar=k_{\rm B}=a=1$
and take $J$ as the unit of energy,
where $a$ is the lattice spacing.


We used the QMC method to draw the ground-state
phase diagram of the mixed-field Ising model on a square lattice.
We adopted the 
Discrete Space Quantum Systems Solver (DSQSS)
library~\cite{motoyama2021},
which implements the directed loop algorithm~\cite{gubernatis2016}.
We chose the periodic-periodic boundary condition
and considered the system sizes $N_{\rm s}=L^2$ with $L\le 32$.
We typically performed $10^5$ Monte Carlo steps to observe
the physical quantities
after discarding $10^5$ Monte Carlo steps for thermalization.
The statistical average was taken over
$64$ independent runs.


\begin{figure}[t]
\includegraphics[width=\columnwidth]{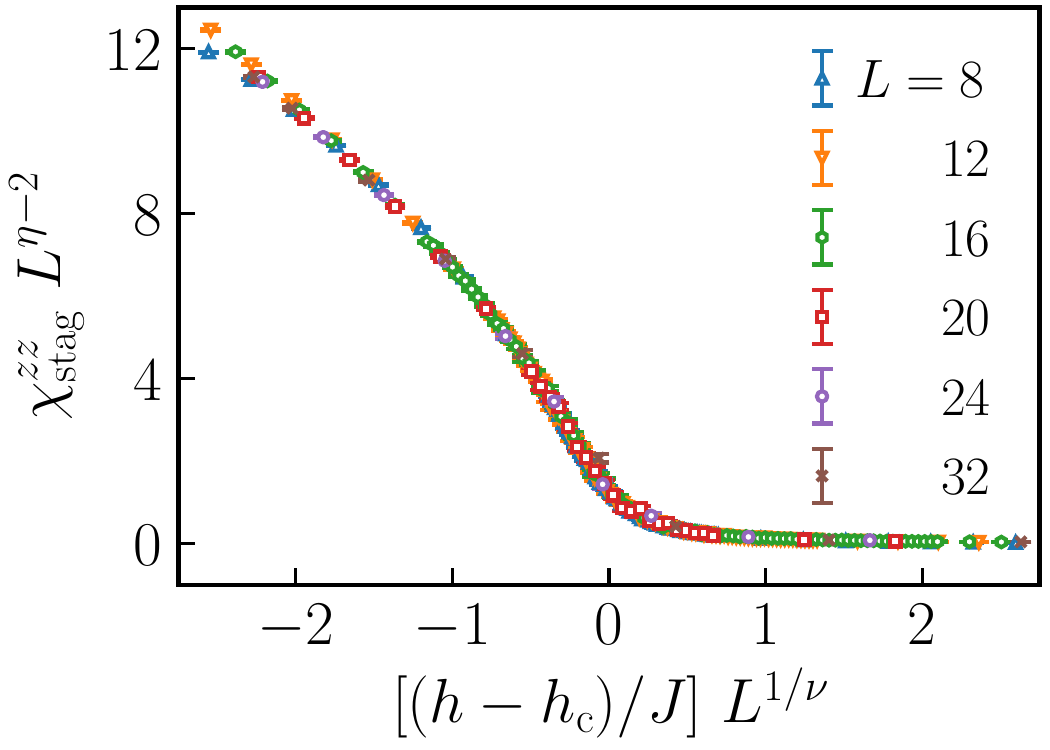}
\caption{(Color online)
Finite-size scaling analysis at $\Gamma/J=0.1$
of the staggered magnetic susceptibility.
Because the transition belongs to the Ising universality class,
we choose
$\eta=0.03631(3)$ and
$\nu=0.62999(5)$~\cite{elshowk2014}.
We also fixed the inverse temperature $\beta J/L=8$ assuming $z=1$.
The transition field is estimated as
$h_{\rm c}/J=2.00426(3)$.}
\label{fig:fss}
\end{figure}

To characterize each phase,
we calculated the staggered magnetic susceptibility~\cite{kato2015},
which is defined as
\begin{align}
 \chi_{\rm stag}^{zz}
 &= \frac{\langle \hat{M}^z(\bm{Q})^2 \rangle}{\beta L^d},
 \quad
 \bm{Q} = (\pi,\pi),
\\
 \text{and}
 \quad
 &
 \hat{M}^z(\bm{q}) = \int_{0}^{\beta} d\tau
 \sum_{j} \hat{S}_j^z(\tau) e^{-i\bm{q}\cdot\bm{r}_j},
\end{align}
where $\bm{r}_j$ is
the real space coordinate at site $j$,
$d(=2)$ is the spatial dimension,
and $\beta$ is the inverse temperature.
When $h$ and $\Gamma$ are
smaller than the critical values,
the ground state is antiferromagnetic,
and the staggered magnetic susceptibility $\chi_{\rm stag}^{zz}$
diverges
as $\beta L^d$
for sufficiently large $L$ and $\beta$.
By contrast,
when $h$ or $\Gamma$ is 
larger than the critical value,
the ground state is disordered,
and $\chi_{\rm stag}^{zz}$ is upper bounded.

We performed finite-size scaling analysis
based on Bayesian scaling analysis~\cite{harada2011}
to determine the phase boundary
between the antiferromagnetic and disordered phases
at zero temperature.
We set the inverse temperature $\beta$
to be proportional to the linear system size $L$
because the dynamical exponent would satisfy $z=1$.
The scaling form~\cite{kato2015}
of $\chi_{\rm stag}^{zz}$ is given as
\begin{equation}
 \chi_{\rm stag}^{zz}
 \sim L^{2-\eta} \mathcal{F}
 (\delta L^{1/\nu}),
\end{equation}
where
$\mathcal{F}$ is a scaling function,
$\eta$ is the anomalous dimension,
and $\nu$ is the correlation length exponent.
The difference of the field from the critical point
is written as
$\delta = (h - h_{\rm c})/J$
$[\delta = (\Gamma - \Gamma_{\rm c})/J]$
for a fixed $\Gamma$ ($h$).
Because the continuous transition for $\Gamma>0$ is expected to
belong to the Ising universality class,
we set the critical exponents to
those in the $(2+1)$D Ising model,
namely,
$\eta=0.03631(3)$ and
$\nu=0.62999(5)$~\cite{elshowk2014}.
The transition field $h_{\rm c}$ ($\Gamma_{\rm c}$)
was estimated for each $\Gamma$ ($h$).
Figure~\ref{fig:fss} shows an example of the finite-size scaling analysis
at $\Gamma/J=0.1$.
The data for different system sizes collapse onto a single curve
at $h\sim h_{\rm c}$.

\begin{figure}[t]
\includegraphics[width=\columnwidth]{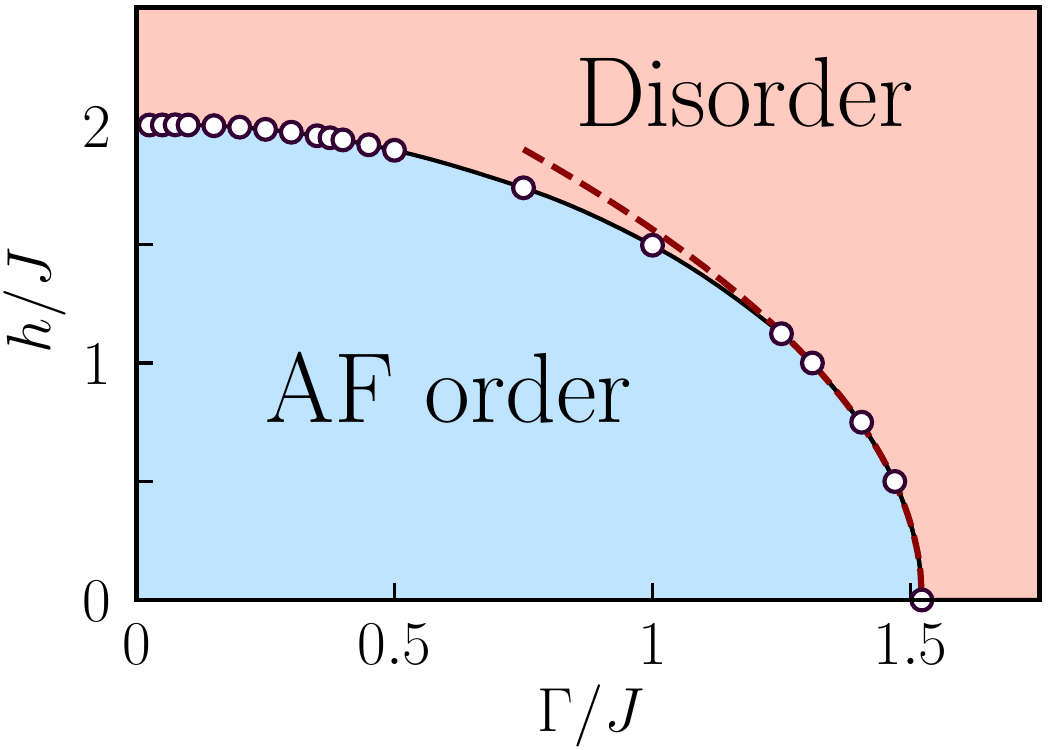}
\caption{(Color online)
Ground-state
phase diagram of the antiferromagnetic Ising model
on a square lattice
with longitudinal and transverse magnetic fields.
The statistical error is smaller than the symbol size.
We
obtained
the transition point at $h=0$
as
$\Gamma_{\rm c}(h=0)=1.5220(4)$,
which is 
consistent with 
previous QMC
studies~\cite{rieger1999,bloete2002}.
The transition longitudinal field $h_{\rm c}$
changes very little and remains $h_{\rm c}\sim 2J$ 
for small $\Gamma$.
The dashed line corresponds to asymptotic behavior
of the phase boundary obtained
by the MF approximation~\cite{ovchinnikov2003},
namely,
$\Gamma_{\rm c}\sim \Gamma_{\rm c}(h=0)
+ 
c_2
h^2 / (2J)$.
Here the 
coefficient obtained by fitting
is given as
$c_2 = -0.427(3)$,
whereas $\Gamma_{\rm c}(h=0)$
is chosen as the transition point in two spatial dimensions.
}
\label{fig:phase_diag}
\end{figure}

We obtain the ground-state phase diagram shown
in Fig.~\ref{fig:phase_diag}.
At $\Gamma=0$, the model becomes classical
and is known to exhibit a first-order transition at
$h_{\rm c}=dJ$~\cite{novotny1986,ovchinnikov2003,sen2000,bonfim2019,
kato2015}.
Our numerical data are consistent with this finding;
for small $\Gamma$,
the transition longitudinal field
changes very little and
remains $h_{\rm c}\sim 2J$.
By contrast,
for $h\sim 0$,
the transition transverse field $\Gamma_{\rm c}$
decreases very sharply.
This observation is similar to the results of the mean-field (MF) approximation~\cite{ovchinnikov2003},
which corresponds to the limit of infinite spatial dimensions,
as well as that in one spatial
dimension~\cite{novotny1986,ovchinnikov2003,sen2000,bonfim2019}.
The phase boundary satisfies
$[\Gamma_{\rm c} - \Gamma_{\rm c}(h=0)]/(dJ)
\sim c_{d} [h/(dJ)]^2$
with
$c_1\sim -0.37$~\cite{novotny1986,ovchinnikov2003,sen2000,bonfim2019}
($c_{\infty}=-0.375$~\cite{ovchinnikov2003})
in one spatial dimension (infinite spatial dimensions).
We found that
$c_2=-0.427(3)$ in two spatial dimensions.
The coefficients $c_1$, $c_2$, and $c_{\infty}$ are approximately $-0.4$.
There seems to be no significant difference between these values
irrespective of the spatial dimensions.

\begin{figure}[t]
\includegraphics[width=\columnwidth]{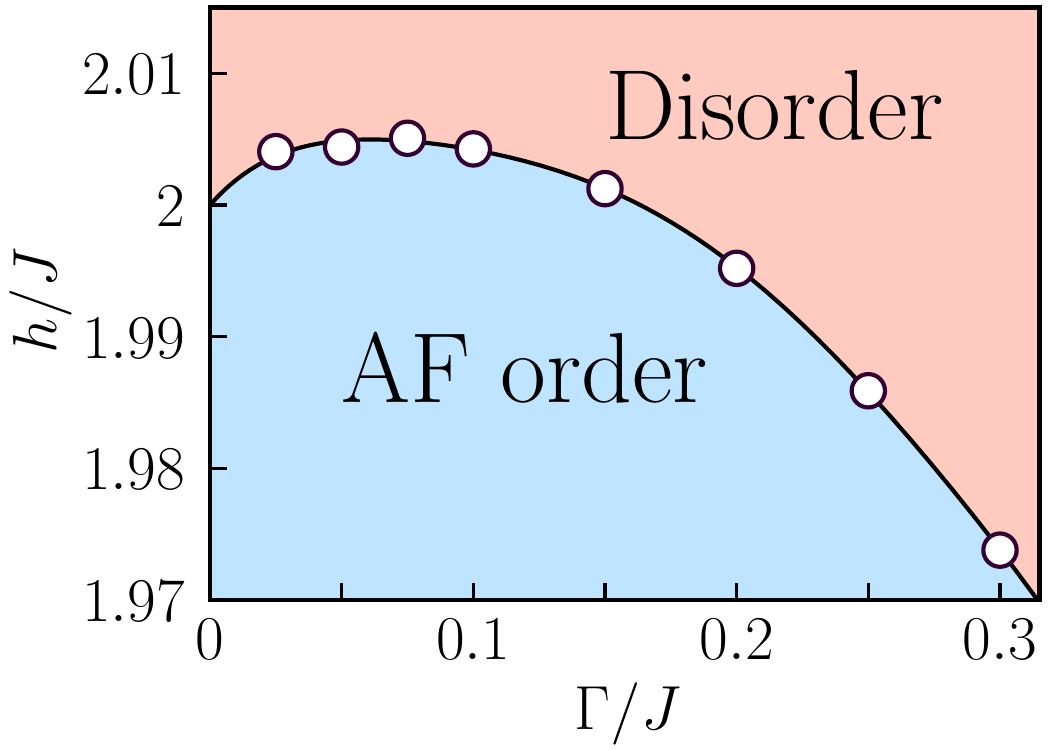}
\caption{(Color online)
Magnified
ground-state
phase diagram near $\Gamma=0$,
where the disordered phase exhibits reentrance.
The statistical error is smaller than the symbol size.
The curve is obtained by fitting the data
to the sextic polynomial for $0\le \Gamma/J \le 0.35$.
For $0\le \Gamma/J\lesssim 0.075$,
the transition longitudinal field $h_{\rm c}$ increases
as the transverse field $\Gamma$ increases.}
\label{fig:phase_diag_magnified}
\end{figure}

Figure~\ref{fig:phase_diag_magnified} shows
the magnified
ground-state
phase diagram for $h\sim 2J$ and $\Gamma\sim 0$.
Remarkably, we found a narrow region
where the disordered phase exhibits reentrance
when the transverse field increases.
Near $\Gamma=0$,
the transition longitudinal field $h_{\rm c}$
increases as the transverse field $\Gamma$ increases.
For $\Gamma \gtrsim 0.075$,
$h_{\rm c}$ starts to decrease as $\Gamma$ increases.
This observation is qualitatively different from
that in one spatial
dimension~\cite{novotny1986,ovchinnikov2003,sen2000,bonfim2019},
where $h_{\rm c}$ decreases monotonically as $\Gamma$ increases.
The transverse field usually destabilizes the antiferromagnetic order
in low spatial dimensions.
Therefore, we expect a monotonic decrease
in the transition field $h_{\rm c}$.
However, this is not the case for the square Ising model.

Unconventional reentrant behavior
has already been observed in the MF
approximation~\cite{ovchinnikov2003},
and our results show that it is maintained even in two spatial dimensions.
Here we briefly review the MF results on the tilt of the spins 
and the variation of the longitudinal transition field
in the presence of a small transverse field.
The MF energy per site
is given as
$E/(dJN_{\rm s}) = [ \cos \theta_1 \cos \theta_2
 - \tilde{\Gamma} (\sin \theta_1 + \sin \theta_2)
 - \tilde{h} (\cos \theta_1 + \cos \theta_2) ]/4$.
Here $dJ$ is taken as the unit of energy,
and the fields are rescaled as
$\tilde{\Gamma}=\Gamma/(dJ)$ and $\tilde{h}=h/(dJ)$.
All spins lie in the $xz$ plane,
and $\theta_1$ ($\theta_2$) is
the angle between the $z$ axis and the spin
at one (the other) sublattice site on a bipartite lattice.
The critical point can be obtained using the stationary condition
$- \sin \theta_i \cos \theta_j
 - \tilde{\Gamma} \cos \theta_i
 + \tilde{h} \sin \theta_i
 = 0$
with $(i,j)=(1,2),(2,1)$.
We would like to obtain
the angle $\theta_{\rm c}(\tilde{\Gamma})$ at the critical point
in the presence of a small transverse field $\tilde{\Gamma}\ll 1$.
Let us define the relative angle between two spins as
$\delta\theta = \theta_2 - \theta_1$.
At the critical point ($\tilde{h}=\tilde{h}_{\rm c}$),
$\delta\theta = 0$, and thus $\theta_1 = \theta_2 = \theta_{\rm c}$.
Therefore, $\theta_{\rm c}$ satisfies
\begin{equation}
 - \sin \theta_{\rm c} \cos \theta_{\rm c}
 - \tilde{\Gamma} \cos \theta_{\rm c}
 + \tilde{h} \sin \theta_{\rm c}
 = 0.
\label{eq:mf_thetac_1}
\end{equation}
Slightly away from the critical point,
we can expand the stationary condition
up to the first order of $\delta\theta\ll 1$.
Replacing $\theta_1$ with $\theta_{\rm c}$,
we obtain
\begin{equation}
 1
 - \tilde{\Gamma} \sin \theta_{\rm c}
 - \tilde{h} \cos \theta_{\rm c}
 = 0.
\label{eq:mf_thetac_2}
\end{equation}
Eliminating $\tilde{h}$ from Eqs.~\eqref{eq:mf_thetac_1}
and \eqref{eq:mf_thetac_2} yields
\begin{equation}
 \sin^3 \theta_{\rm c} = \tilde{\Gamma}.
\label{eq:mf_theta3_g}
\end{equation}
Without the transverse field $\tilde{\Gamma}$,
the spins align in the $z$ direction,
and the angle at $\tilde{h}=\tilde{h}_{\rm c}+0$
is $\theta_{\rm c}=0$.
For small $\tilde{\Gamma}$,
Eq.~\eqref{eq:mf_theta3_g} suggests that
the tilting angle grows more rapidly than it does in a linear field
($\theta_{\rm c}\sim \tilde{\Gamma}^{1/3}$).
The $z$ component ($x$ component)
of the spin becomes significantly smaller (larger).
To compensate for the smaller $z$ component,
the longitudinal field should be sufficiently large to cause
the transition; that is,
$\tilde{h}_{\rm c}(\tilde{\Gamma}\ll 1) >
\tilde{h}_{\rm c}(\tilde{\Gamma}=0)$.
Indeed,
from Eqs.~\eqref{eq:mf_thetac_2} and \eqref{eq:mf_theta3_g},
the transition field satisfies
$\tilde{h}_{\rm c}
=(1-\tilde{\Gamma}^{2/3})^{1/2}(1+\tilde{\Gamma}^{2/3})
\sim 1 + \tilde{\Gamma}^{2/3}/2$,
which clearly indicates reentrance.

As we see below,
subtle competition between the MF mechanism and quantum
fluctuations determines whether reentrance occurs.
The ground state at $(\Gamma,h)=(0,dJ)$ is macroscopically
degenerate~\cite{domb1960,ovchinnikov2003,sen2000,
kato2015,wurtz2020,lajko2021}.
The first-order transition at this point
becomes a continuous one
in the presence of an infinitesimally small $\Gamma$.
The shape of the phase boundary near
$\Gamma=0$ is susceptible to quantum fluctuations,
which can vary depending on
the spatial dimensions of the system.
In one spatial dimension,
strong quantum fluctuations destabilize the
antiferromagnetic order.
The transition longitudinal field behaves as
$h_{\rm c} \sim J + c'_{1} \Gamma$
with a negative coefficient
$c'_{1}\sim -0.68(4)$~\cite{igloi1989,ovchinnikov2003,lin2017,lajko2020a}.
By contrast,
in the MF approximation,
which 
can be regarded as
the limit of infinite spatial dimensions, the
antiferromagnetic order is favored.
The transition obeys the relation
$h_{\rm c} \sim J + c'_{\infty} \Gamma^{2/3}$
with a positive coefficient
$c'_{\infty} = 0.5$~\cite{ovchinnikov2003}.
On a square lattice,
we found numerically that the transition satisfies
$h_{\rm c} \sim 2J + c'_{2} \Gamma$
with a very small positive coefficient,
$c'_{2}\sim 0.16$.
Here we assume that the longitudinal transition point
is a linear function of the transverse field for $\Gamma\sim 0$.
This behavior is predicted by perturbation theory
in one spatial dimension~\cite{ovchinnikov2003},
although 
it does not have to occur in general.
The monotonic behavior of the coefficients
($c'_1 < c'_2 < c'_{\infty}$) suggests that
the shape of the phase boundary in two spatial dimensions
is intermediate between those in one and infinite spatial dimensions.


In conclusion,
we studied the antiferromagnetic Ising model
on a square lattice
with longitudinal and
transverse magnetic fields
using the QMC method.
We determined the phase boundary between
the antiferromagnetic and disordered phases
by finite-size scaling analysis
and found that the critical field $h_{\rm c}$
changes very little for small $\Gamma$.
We also found
a narrow
region where the disordered phase exhibits reentrance
near the point $(\Gamma,h)\sim (0,2J)$.
By comparing our result with those of previous studies,
we found that
the shape of the phase boundary in two spatial dimensions
is intermediate between 
those in one and infinite spatial dimensions.
Reentrant behavior,
which occurs in infinite spatial dimensions,
seems to emerge in the rather low dimensionality of two spatial dimensions.

Our phase diagram would be helpful for analog quantum simulation 
of Rydberg systems.
Although it is challenging to detect the narrow reentrant
region on a square lattice,
a nearly intact transition field $h_{\rm c}$ as a function
of $\Gamma$ may be observed for $\Gamma\sim 0$.
This behavior can be measured to confirm the accuracy
of quantum simulations.

In three spatial dimensions,
the MF critical exponent will be exact
because the value $d+z=4$ reaches the upper critical dimension.
We may observe a broader reentrant region.
It is of great interest to investigate how the phase boundary
is modified on a cubic lattice.
This topic remains a subject of future study.
In addition,
atoms in Rydberg states are governed by
the van der Waals interaction.
It is also interesting to study the Ising model with more realistic
long-range interaction~\cite{samajdar2020} and related models in
geometrically frustrated
lattices~\cite{kellerman2019,wang2021,samajdar2021}.
This topic also remains a subject of future study.

 
\acknowledgments

The authors acknowledge fruitful discussions with
T.\ Uno.
This work was financially supported by JSPS KAKENHI
(Grants Nos.\ 18K03492 and 18H05228),
by JST CREST (Grant No.\ JPMJCR1673),
and by MEXT Q-LEAP (Grant No.\ JPMXS0118069021).
The numerical computations were performed on computers at
the Yukawa Institute Computer Facility and on computers at
the Supercomputer Center, the Institute for Solid State Physics,
the University of Tokyo.


\bibliographystyle{apsrev4-2}
\input{main.bbl.back}

\onecolumngrid


\end{document}

%% file: main.bbl.back
%